\documentclass[preprint]{aastex}
\usepackage{amsmath, amsthm}
\usepackage{lscape}
\usepackage{graphics}
\usepackage{epsfig}
\usepackage{graphicx}
\usepackage{rotate}
\usepackage{color}

\def\ARAA{{\it Annual Rev. of Astron. \& Astrophys.} }
\def\ApJ{{\it Astrophys. J.} }

\def\AA{{\it Astron. \& Astroph.} }

\def\MNRAS{{\it Month. Not. Roy. Astr. Soc.} }
\def\Nature{{\it Nature} }

\def\PRD{{\it Phys. Rev.} {\bf D} }
\def\PRL{{\it Phys. Rev. Letters} }

\def\simle{\lower 2pt \hbox {$\buildrel < \over {\scriptstyle \sim }$}}
\def\simge{\lower 2pt \hbox {$\buildrel > \over {\scriptstyle \sim }$}}

\shorttitle{WMAP haze and massive star explosions}
\shortauthors{Biermann et al.}

\begin{document}


\title{The WMAP haze from the Galactic Center region due to massive star 
explosions and a reduced cosmic ray scale height}


\author{Peter L.~Biermann\altaffilmark{1,2,3,4}}
\affil{MPI for Radioastronomy, Bonn, Germany}
\email{plbiermann@mpifr.mpg.de}
\author{Julia K.~Becker}
\affil{Ruhr-Universit\"at Bochum, Fakult\"at f\"ur Physik \&
  Astronomie, Theoretische Physik IV, D-44780 Bochum, Germany}
\author{Gabriel Caceres}
\affil{Dept of Astron. \& Astroph., Penn State U., University Park, PA, USA}
\author{Athina Meli}
\affil{ECAP, Physik. Inst. Univ. Erlangen-N{\"u}rnberg, Germany}
\author{Eun-Suk Seo}
\affil{Dept. of Physics, Univ. of Maryland, College Park, MD, USA}
\and
\author{Todor Stanev}
\affil{Bartol Research Inst., Univ. of Delaware, Newark, DE, USA}
 
\altaffiltext{1}{Dept. of Phys. \& Astron., Univ. of Bonn, Germany }
\altaffiltext{2}{Dept. of Phys. \& Astr., Univ. of Alabama, Tuscaloosa, AL, USA}
\altaffiltext{3}{Dept. of Phys., Univ. of Alabama at Huntsville, AL, USA}
\altaffiltext{4}{FZ Karlsruhe, and Phys. Dept., Univ. Karlsruhe, Germany}


\begin{abstract}
One important prediction of acceleration of particles in the supernova 
caused shock in the magnetic wind of exploding Wolf Rayet and Red Super 
Giant stars is the production of an energetic particle component with a 
$E^{-2}$ spectrum, at a level of a few percent in flux at injection. 
After allowing for transport effects, so steepening the spectrum to
$E^{-7/3}$, this component of electrons produces electromagnetic
radiation and readily explains the WMAP haze from the Galactic Center region in
spectrum, intensity and radial profile.
This requires the diffusion time scale for cosmic rays 
in the Galactic Center region to be much shorter than in the Solar 
neighborhood: the energy for cosmic ray electrons at the transition 
between diffusion dominance and loss dominance is shifted to 
considerably higher particle energy. We predict that more precise 
observations will find a radio spectrum of $\nu^{-2/3}$, at higher 
frequencies $\nu^{-1}$, and at yet higher frequencies finally $\nu^{-3/2}$.
\end{abstract}



\section{Introduction}

Scanning all sky the WMAP satellite discovered a haze in the region of 
the Galactic center, with radio frequencies up to near 100 GHz, and a 
relatively flat spectrum (Finkbeiner 2004a, 2004b; Hooper et al. 2007; 
Dobler \& Finkbeiner 2008; Caceres 2009). There are a number of possible 
explanations for this haze, as discussed in these papers, such as 
annihilation of dark matter, and also attributing the haze to various 
stars (Bandyopadhyay et al. 2009); however, those authors conclude their 
specific proposal fails. The radio spectrum of this haze is very much 
flatter at such frequencies than predicted on the basis of cosmic ray 
electron data (Wiebel-Sooth \& Biermann 1999), and so there is an 
apparent conflict. Here we estimate whether this extra emission can be 
attributed to the polar cap component of cosmic rays, emanating from 
massive star explosions, following reasoning in earlier work (Biermann 
1993, Biermann \& Cassinelli 1993, Stanev et al. 1993, Biermann et al. 
2001, 2009). The earlier prediction had been verified in Biermann et al. 
(2009), finally definitively showing the existence of the cosmic ray 
polar cap component. In the following we use cgs units.

We find, that the polar cap component can explain the data. It requires 
the magnetic turbulence in the Galactic center region to be much 
stronger than in the Solar neighborhood (see Aharonian et al. 2006, 
Becker et al. 2009), so that first the diffusion scale height for cosmic 
rays is reduced, and so second the transition between the 
diffusion-dominated regime and the loss-dominated regime is shifted to 
much higher energy than in the Solar neighborhood.

\section{Cosmic ray transport and loss across the Galaxy}

Cosmic rays are injected from massive star explosions, with predicted 
spectra of $E^{-2.42 \pm 0.04}$ from ISM-SNe (Biermann \& Strom 1993), 
$E^{-7/3 - 0.02 \pm 0.02}$ from wind-SNe (Biermann 1993, Biermann \& 
Cassinelli 1993), with a polar cap component at a few percent level and 
a spectrum of $E^{-2}$ (Stanev et al. 1993). All these spectra are 
steepened by diffusive losses by 1/3, so become $E^{-2.75 \pm 0.04}$, 
$E^{-8/3 - 0.02 \pm 0.02}$, and $E^{-7/3}$ (Stanev et al. 1993). The 
polar cap component is limited by spatial constraints to a maximum 
energy of order PeV (for Hydrogen). The polar cap component is slower in 
its acceleration, and so has more time to interact and produce 
secondaries (Jokipii 1987, Meli \& Biermann 2006), so naturally 
explaining quantitatively the ATIC, H.E.S.S., Pamela, and Fermi results 
for cosmic ray electrons and positrons (Pamela: 
Adriani et al. 2009; H.E.S.S.: Aharonian et al. 2008; ATIC: Chang 
et al. 2008) with such a concept (Biermann et al. 2009). H.E.S.S. 
(Aharonian et al., 2009) and Fermi (Abdo et al., 2009) gave further 
quantitative confirmation of the predictions.

In the Solar neighborhood we have a competition between diffusive 
losses, running with $E^{-1/3}$, and synchrotron and inverse Compton 
losses, running with $E^{-1}$. The transition for cosmic ray electrons 
is at about 20 GeV, so a Lorentz factor of about $\gamma_e \, = \, 
10^{4.6}$. At a similar energy the wind-SN-electrons may begin to 
dominate over the ISM-SN-electrons, so shifting the spectrum from 
$E^{-2.75}$ to $E^{-8/3}$; this is visible in data comparing radio 
spectra with cosmic electron spectra (Wiebel-Sooth \& Biermann 1999). 
Above this energy the spectrum is $E^{-10/3}$, before the polar cap 
component rises up to dominate at $E^{-3}$, at a transition energy of 
about 40 GeV, corresponding to a Lorentz factor of $\gamma_e \, = \, 
10^{4.9}$. Since at that energy the ISM-SN-cosmic rays still contribute 
appreciable partial flux, the straight transition to polar cap dominance 
for wind-SN-cosmic ray particles only may be at slightly lower energy 
even, such as at $\gamma_e \, \simeq \,
10^{4.4}$, or even less. We note in passing, that at 
these energies the positrons do not contribute a strong partial flux 
(see the graphs in Biermann et al. 2009).

The transition between the cosmic ray spectrum contributed from most of 
$4 \pi$ of the sphere of a shock caused by a massive star explosion, and 
racing through the wind, is just a function of stellar physics (Langer 
\& Heger 1999; Heger et al. 2003); the ratio between ISM-SNe, so 
slightly lower mass stars, and wind-SNe, is a function of star formation 
activity; in a starburst, temporarily more massive stars are born (e.g. 
Biermann \& Fricke 1977, Kronberg et al. 1985), and so their 
contribution to cosmic ray fluxes, including the polar cap component, is 
temporarily stronger relative to that of the ISM-SNe. In the Galactic 
Center region we may have a top-heavy mass distribution of stars, so 
more very massive stars (Bartko et al. 2009).

However, the transition between the diffusion regime and the loss 
regime, very strongly depends on the magnetic turbulence: The stronger 
the turbulence, the faster cosmic ray transport, and the higher the 
transition energy between the two regimes (Aharonian et al. 2006; Becker 
et al. 2009). In the Galactic Center region star formation peaks for the 
entire Galaxy, and so it is altogether very plausible, that the 
transport by diffusive leakage is much faster than in the Solar 
neighborhood; in fact it can be analytically shown, that the scale 
height for the diffusive regime scales inversely with the strength of 
the magnetic turbulence (Becker et al. 2009). It follows that the 
diffusive time scale, estimated for the Solar neighborhood at near 
$10^{7}$ years, is very much shorter in the Galactic Center region.

The WMAP haze has been observed at radio frequencies between 20 and near 
100 GHz, and at about 10 $\mu$Gau{\ss} magnetic field strength 
(Berkhuijsen 1994, in Beck et al. 1994; Ferri{\`e}re 2009) this implies a 
range of cosmic ray electron Lorentz factors of $\gamma_e$ from 
$10^{4.35}$ to $10^{4.7}$. The precise spectrum is very difficult to 
determine due to the background subtraction errors; the data suggest a 
hardening by 0.3 in radio spectral index with respect to the normal 
radio emission with spectral index -0.88 (corresponding to the observed 
cosmic ray proton spectrum, consistent with the inferred cosmic ray 
electron spectrum below about 10 GeV (Wiebel-Sooth \& Biermann 1999; 
Hooper et al. 2007); this entails that the haze spectrum is somewhere 
near $\nu^{-0.58}$, but with relatively large error bars. The data 
suggest that the flat component extends to about 100 GHZ. This then 
implies that the diffusive transport time scale in the Galactic Center 
region is at least 5 times shorter than in the Solar neighborhood (for 
GeV particles).

To explain the WMAP haze all we require is, a) that at these cosmic ray 
electron energies we are still in the diffusive regime of cosmic ray 
transport, and b) the polar cap component has already started to come 
up. This suggests a cosmic ray electron spectrum in
the relevant energy range of of $E^{-7/3}$, and so 
correspondingly a radio spectrum at high frequencies of $\nu^{-2/3}$. 
Following the reasoning in Biermann et al. (2009) we can then predict 
the transition to the loss dominated regime, giving a radio spectrum at 
higher frequencies of $\nu^{-1}$, and finally the spectrum resulting 
from a limited spatial reach, $\nu^{-3/2}$. The Planck satellite may be 
able to test these clear predictions. Due to a very strong dependence on 
the diffusion scale height, it is difficult to pinpoint the exact radio 
frequencies of the transitions between the different regimes.

\subsection{Cutoff frequencies}

Can we predict the frequencies of the various transitions between the 
three spectral regimes?

The loss time of cosmic ray electrons is

\begin{equation}
\tau_{syn} \; = \; \frac{6 \pi m_e c}{\sigma_T \gamma_e B^2 \epsilon_{IC}}
\end{equation}

where $\sigma_T$ is the Thompson cross section, $\gamma_e$ is the 
Lorentz factor of the emitting electrons (or positrons), $B^2/8 \pi$ is 
the magnetic field energy density, while $\epsilon_{IC}$ is the 
enhancement of the losses due to inverse Compton interaction with the 
ambient photon field.

The diffusive loss time of cosmic ray electrons is

\begin{equation}
\tau_{diff} \; = \; \tau_0 \, {\gamma_{e, 3.3}}^{-1/3}
\end{equation}

where $\tau_{0} \simeq 10^{7} \, {\rm yrs}$, and $\gamma_{e, 3.3}$ is 
the electron Lorentz factor at the energy corresponding to the proton 
rest mass, where we can estimate the time scale from cosmic ray data in 
the Solar neighborhood.

For a given cosmic ray electron energy the scaling of the time scale for 
diffusion is

\begin{equation}
\tau_0 \; \sim \; H_{CR}^{2} \, B \, b
\end{equation}

where $H_{CR}$ is the diffusion scale height, about 1 - 2 kpc in the 
Solar neighborhood; $B$ is again the magnetic field strength, and $b$ is 
the ratio between the turbulent energy in magnetic field fluctuations, 
and the overall magnetic field energy density.

The transition between losses via synchrotron and Inverse Compton losses 
on one side, and diffusive losses on the other is at about 10 - 20 GeV 
in the Solar neighborhood, and at a much higher energy in the Galactic 
Center region, we claim.

Inserting numbers gives

\begin{equation}
\tau_{syn} \; = \; 10^{7.7} \; {\rm yrs} \; B_{-5}^{-2} \, \gamma_{e, 
3.3}^{-1}
\end{equation}

using $10^{-5}$ Gauss as scale for the magnetic field.

The transition energy $\gamma_{e,3.3, \star}$ is then given by

\begin{equation}
\gamma_{e, 3.3, \star}^2 \; = \; 10^{2.1} \, B_{-5}^{-6} \, 
\left(\frac{H_{CR,GC}}{H_{CR, sol}}\right)^{-6} \, 
\left(\frac{B_{GC}}{B_{sol}}\right)^{-3} \, 
\left(\frac{b_{GC}}{b_{sol}}\right)^{-3} \, \left(\frac{\epsilon_{IC, 
GC}}{\epsilon_{IC, sol}}\right)^{-3}
\end{equation}

which we can insert into the expression for the cut-off frequency.

Using 10 $\mu$Gau{\ss} for the magnetic field at the Galactic Center, as 
well as a factor of 2 between the magnetic field there and in the Solar 
neighborhood, and allowing a frequency of 100 GHz or higher for the haze 
then requires

\begin{equation}
\left(\frac{H_{CR,GC}}{H_{CR, sol}}\right) \, \; < \; 0.5 \, 
\left(\frac{b_{GC}}{b_{sol}}\right)^{-1/2} \, \left(\frac{\epsilon_{IC, 
GC}}{\epsilon_{IC, sol}}\right)^{-1/2}
\end{equation}

Of course, this could be much less than the limit. The two factors on 
the right hand side, the enhancement in magnetic turbulence at the 
Galactic Center relative to the Solar neighborhood, and the enhancement 
in radiation fields, are both expected to be much larger than unity. 
Therefore the scale height has to be much smaller in the Galactic Center 
region.

However, this also illustrates, that the dependence on the precise 
parameters is so strong, that we may well have to determine the cutoff 
frequencies from observations, and from that constrain the main 
parameters, with the diffusive scale-height $H_{CR}$ the most critical. 
Decreasing the scale height by a relatively small factor can increase 
the critical turnoff frequency enormously.

\subsection{Flux}

The observed spectrum of energetic electrons from the polar cap component is

\begin{equation}
\frac{d N_e}{d \gamma_e} \; = \; 10^{+4.75} \gamma_e^{-3} \rm{cm^{-2} 
sr^{-1} s^{-1}}
\end{equation}

Going down from the loss limit (Kardashev 1962) to the transition energy 
for the diffusion limit gives

\begin{equation}
\frac{d N_e}{d \gamma_e} \; = \; 10^{+1.75} \gamma_e^{-7/3} \rm{cm^{-2} 
sr^{-1} s^{-1}}
\end{equation}

This implies a density of particles of

\begin{equation}
C_e \; = \; 10^{-7.65} \gamma_e^{-7/3} {\rm cm^{-3}}
\end{equation}

This has been inferred for the Solar neighborhood; in the Galactic 
Center region the flux is likely to be higher by about the magnetic 
field energy density, so by about 4 (Berkhuijsen, in Beck et al. 1996). 
So we finally obtain an estimate of the polar cap component electrons in 
the Galactic Center region of

\begin{equation}
C_e \; = \; 10^{-7.05} \gamma_e^{-7/3} {\rm cm^{-3}}
\end{equation}

Taking a reference volume corresponding to $\theta_{\star} \, = \, 0.1$ 
rad, so about 6 degrees, the observations (e.g. Hooper et al. 2007) give 
about 4 kJy sr$^{-1}$ emission. Here we take just one point on their 
curves, and estimate the flux density from their figures, at that 
angular distance, and discuss the radial variation separately. This has 
to be matched with

\begin{equation}
S_{haze} \; = \; \frac{Vol}{4 \pi D_{GC}^{2}} \, 3 \cdot 10^{-18} \, C_e 
\, B^{5/3} \, \nu^{-2/3} \frac{1}{\Delta \Omega}
\end{equation}

where $\Delta \omega$ is the solid angle of this emission, as seen from 
us, and using $C_e \, \gamma_e^{-7/3}$ as the energetic electron 
spectrum. As ${Vol} \, \sim \, \theta_{\star}^{3}$ and $\Delta \Omega \, 
\sim \, \theta_{\star}^{2}$, the dependence of $S_{haze} \, \sim \, 
\theta_{\star}$, so linear. The volume of this limited region near the 
Galactic Center as delineated above is about $10^{64.8}$~cm$^3$, and 
the distance is about 7.5 kpc, so that $4 \pi D_{GC}^{2} \, \simeq \, 
10^{45.9} \, {\rm cm^{2}}$. Here we then obtain at about 100 GHz, over 
$\Delta \Omega \simeq 10^{-2}$ sr

\begin{equation}
S_{haze} \; = \; 10^{+3.55} \, {\rm Jy \, sr^{-1}}
\end{equation}

to compare with $10^{3.6} \; {\rm Jy\, sr^{-1}}$, so a reasonable match, 
considering the uncertainties clear from the derivation.

\subsection{Radial profile}

Hooper et al. (2007) give an approximate flux density profile of about 
$s^{-1.4}$, with $s$ being the perpendicular distance to the center of 
the disk. Therefore, there is an important question: How far out above 
the Galactic plane can we predict the flat component to be visible? 
Writing the losses of a charged particles moving along with the Galactic 
wind (Westmeier et al. 2005, Breitschwerdt 2008, Everett et al. 2008), 
referring the time of transition from the diffusive regime to the 
convective regime

\begin{equation}
\frac{d \gamma_{e}}{d t} \; = \; - \frac{1}{2} \, \frac{\gamma_{e}}{t} 
\, - \, \frac{\gamma_{e}^{2}}{\tau_{syn, 0}} \, \left(\frac{t_0}{t} 
\right)^{2}.
\end{equation}

Here we have used magnetic moment conservation, the asymptotic Parker 
regime of the wind, where the magnetic field runs as $1/r$ (Parker 
1958), and assume a conical cut from a spherical configuration. For more 
cylindrical or more fountain-like geometries the factors $1/2$ and 
$1/t^{2}$ get modified. We assume, that the wind is quasi stationary, 
and so that time $t$ is equivalent to distance, using here polar 
coordinates.

There are two extreme solution: as long as the first term dominates the 
solution is

\begin{equation}
\gamma_{e} \; \sim \; \left(\frac{t}{\tau_{0}}\right)^{-1/2}.
\end{equation}

If the second term were to dominate, this condition implies

\begin{equation}
\gamma_{e} \; \simge \; \frac{1}{2} \, \frac{\tau_{syn, 0}}{t_{0}} \, 
\frac{t}{t_0}
\end{equation}

so the transition energy increases with time; setting our attention at 
the cutoff edge at the start of the wind, where losses just begin to 
become important, this expression shows, that adiabatic losses always 
win, as the transition energy to synchrotron and Inverse Compton losses 
goes up with time, as we follow the particle population out along with 
the wind.
Therefore the emission from the flatter component of the electron 
population stays always dominant, as the particles move out through the 
wind.

The derivation above also implies, that the momentum of a particle runs 
as $p \, \sim \, r^{-1/2}$, a spectrum $N_{CR}(p) d p$ decreases as 
$r^{-2/3}$, and so the emission integrated along the line of sight 
through the cone runs as $r^{-4/3}$, close to what is observed. 
Therefore this simple model lets us understand vertical profile, and 
spectrum.

Last we estimate how many cosmic ray electrons participate in producing 
the WMAP haze: We take the density from above in the nonthermal 
spectrum, integrate from the minimum energy of cosmic ray electrons (see 
Biermann et al. 2009), a solid angle of $10^{-2}$ sr consistent with the 
earlier crude numbers, the distance to the Galactic Center of about 8 
kpc, and the flow velocity of about 1000 km/s, a small multiple of the 
escape speed.
 We assume, that the area around the Galactic Center 
participating in the flow is reasonably well described by the same 
length scale as what we observe on the sky, and so arrive at a crude 
estimate of $10^{41}$ s$^{-1}$ of polar cap cosmic ray electrons 
produced. The normal steeper spectrum cosmic ray electrons are a factor 
of order 30 (the inverse of a few percent) larger, estimated over the
same area of the Galactic disk around the Galactic Center, so about $10^{42.5}$ s$^{-1}$, and correspondingly larger for 
a larger surface area, like the 3 kpc ring region, where most of the 
star formation happens in the Galaxy; for such a larger region the 
production rate is then would be another factor of about $10^{1.6}$ 
larger, so about $10^{44.1}$ s$^{-1}$. As the production of secondaries 
rises towards lower energy, the cosmic ray positron production may 
approach $10^{43.5}$ s$^{-1}$. The ensuing annihilation rate may explain 
the 511 keV emission line in the Galactic Center (e.g. Weidenspointner, 
et al. 2008). Obviously, this energy supply in these cosmic ray 
electrons is only a small fraction of the entire cosmic ray energy 
output even in the inner Galaxy. We will consider secondary production 
and the 511 keV emission line elsewhere in detail.

\section{Conclusions}

Following earlier reasoning we propose that massive star explosions give 
rise to a cosmic ray component from their polar cap, which has a flatter 
spectrum, but also slower acceleration times, and can so explain the 
WMAP haze, a zone of flatter radio spectrum at high frequencies around 
the Galactic Center. The condition is that in the Galactic Center region 
the transport of cosmic ray particles is considerably faster, so that 
the transition between diffusive losses and synchrotron/inverse Compton 
losses is shifted to higher particle energy relative to the Solar 
neighborhood. It can be shown (Becker et al. 2009) that the high star 
formation rate per area in that region of the Galaxy leads to a short 
transport time.

This is now further support for the concept of a polar cap component in 
cosmic rays from very massive star explosions (Biermann 1993, Stanev et 
al. 1993, Biermann et al. 2009). This also adds support for the 
magneto-rotational mechanism for massive star explosions (Kardashev 
1964, Bisnovatyi-Kogan 1970, Ardeljan et al. 2000, Moiseenko et al. 
2003), since our argument requires that the particle energy at
which this component becomes relevant is very nearly the same for all very massive stars (Biermann 
1993).

We argue, that in the centers of galaxies and starburst galaxies the 
cosmic ray diffusion loss time scales can become considerably shorter 
than in the Solar neighborhood. This will help understand starburst 
galaxies such as M82 (Kronberg et al. 1985). So we predict, that in the 
central parts of other galaxies as well as star burst galaxies the scale 
height is also considerably reduced compared to the Solar neighborhood.

Our main prediction is the spectrum of the haze as a sequence of 
spectral components of $\nu^{-2/3}$, $\nu^{-1}$, and $\nu^{-3/2}$.

\acknowledgments
Discussions (PLB) with Carlos Frenk are gratefully acknowledged.
Support for work with PLB has come from the AUGER membership and theory 
grant 05 CU 5PD 1/2 via DESY/BMBF and VIHKOS. Support for ESS comes from 
NASA grant NNX09AC14G and for TS comes from DOE grant
UD-FG02-91ER40626. JKB is supported by the Research Department of
Plasmas with Complex Interactions (Bochum).

\end{document}